\documentclass[longauth]{aa}

\usepackage{graphicx}

\newcommand{\etal}{\hbox{et al.} }

\begin{document}

\title{3.9 day orbital modulation in the TeV $\gamma$-ray flux and spectrum from the X-ray binary LS~5039} 
\titlerunning{Orbital modulation in LS~5039}

\author{F. Aharonian\inst{1} \and
  A.G.~Akhperjanian \inst{2} \and
  A.R.~Bazer-Bachi \inst{3} \and
  M.~Beilicke \inst{4} \and
  W.~Benbow \inst{1} \and
  D.~Berge \inst{1} \and
  K.~Bernl\"ohr \inst{1,5} \and
  C.~Boisson \inst{6} \and
  O.~Bolz \inst{1} \and
  V.~Borrel \inst{3}
% ...
  \\[3mm] \mailname{ denauroi@in2p3.fr, growell@physics.adelaide.edu.au}
}

\institute{
  Max-Planck-Institut f\"ur Kernphysik, P.O. Box 103980, D 69029
  Heidelberg, Germany
\and
  Yerevan Physics Institute, 2 Alikhanian Brothers St., 375036 Yerevan,
  Armenia
\and
  Centre d'Etude Spatiale des Rayonnements, CNRS/UPS, 9 av. du Colonel Roche, BP
  4346, F-31029 Toulouse Cedex 4, France
\and
  Universit\"at Hamburg, Institut f\"ur Experimentalphysik, Luruper Chaussee
  149, D 22761 Hamburg, Germany
\and
  Institut f\"ur Physik, Humboldt-Universit\"at zu Berlin, Newtonstr. 15,
  D 12489 Berlin, Germany
\and
  LUTH, UMR 8102 du CNRS, Observatoire de Paris, Section de Meudon, F-92195 Meudon Cedex,
  France
% ...
}

\authorrunning{Aharonian et al.}

\date{Received / Accepted}


\abstract
{}
{ LS~5039 is a High Mass X-ray Binary (HMXRB) comprising a compact object in an eccentric 3.9~day orbit around a massive O6.5V star.
  Observations at energies above 0.1~TeV (10$^{11}$~eV) by the High Energy Stereoscopic System (HESS) in 2004 revealed that LS~5039 is a source
  of Very High Energy (VHE) $\gamma$-rays and hence, is able to accelerate particles to multi-TeV energies. Deeper observations by HESS were carried out
  in 2005 in an effort to probe further the high energy astrophysics taking place. In particular, we have searched for orbital modulation of the
  VHE $\gamma$-ray flux, which if detected, would yield new information about the complex variation in $\gamma$-ray absorption and production within X-ray binary systems.}
{ Observations at energies above 0.1~TeV (10$^{11}$~eV), were carried out with the High Energy Stereoscopic System (HESS) 
  of Cherenkov Telescopes in 2005. A timing analysis was performed on the dataset employing the Lomb-Scargle and Normalised Rayleigh statistics,
  and orbital phase-resolved energy spectra were obtained.}
{ The timing analysis reveals a highly significant (post-trial chance probability $<10^{-15}$) peak 
  in the TeV emission periodogram at a frequency matching that of the 3.9~day orbital motion of the compact object around the massive stellar companion. 
  This is the first time in $\gamma$-ray astronomy that orbital 
  modulation has been observed, and periodicity clearly established using ground-based $\gamma$-ray detectors.
  The $\gamma$-ray emission is largely confined to half of the orbit, peaking around the inferior conjunction epoch of the compact object.
  Around this epoch, there is also a hardening of the energy spectrum in the energy range between 0.2~TeV and a few~TeV.}
{ The $\gamma$-ray flux vs. orbital phase profile suggests the presence of $\gamma$-ray absorption via pair production, 
  {{\bf which would imply}} that a large fraction of the $\gamma$-ray production region is situated within $\sim$1~AU of the compact object.
  This source size constraint can be compared to the collimated outflows or jets observed in LS~5039 resolved down to scales of a few~AU.
  The spectral hardening is however not explained exclusively by the absorption effect, indicating that other effects are present, perhaps related to 
  the $\gamma$-ray production mechanism(s).
  If the $\gamma$-ray emission arises from accelerated electrons, the hardening may arise from variations with phase in the maximum electron energies, the dominant
  radiative mechanism, and/or the angular dependence in the inverse-Compton scattering cross-section. Overall, these results provide new insights into the 
  competing $\gamma$-ray absorption and production processes in X-ray binaries.}

\maketitle

\section{Introduction}

X-ray binaries (XRBs) comprise a compact object such as a neutron star or black hole in orbit with a companion star. They are
one of several types of astrophysical system that can provide a periodic environment for the acceleration of 
particles and subsequent production of radiation. 
Modulation of this radiation, linked to the orbital motion of the binary system, provides key insights into the nature and location of 
particle acceleration and emission processes. 
While such modulation is often found in XRBs up to hard X-ray energies (Lewin \cite{Lewin:1},  Wen \etal \cite{Wen:1}), until now it has not 
been established in any astrophysical source at $\gamma$-ray energies.
LS~5039 (distance $d\sim$2.5~kpc) is a HMXRB comprising a compact object in a $\sim$3.9~day orbit around a 
massive O6.5V star (Casares \etal \cite{Casares:1}). Persistent radio outflows (observed with extension in the range 
2 to $\sim$1000~AU) are attributed to a mildly relativistic ($v\sim 0.2c$) jet (Paredes \etal \cite{Paredes:1,Paredes:2}), 
which would place LS~5039 in the {\em microquasar} class. Microquasars are Galactic, 
scaled-down versions of Active Galactic Nuclei (AGN) (Mirabel \cite{Mirabel:1}), and are a sub-class of XRBs.
The detection of {\bf radio (Mart\'i \etal \cite{Marti:1}, Rib\'o \etal \cite{Ribo:2}) and X-ray (Bosch-Ramon \etal \cite{Valenti:2}) emission
and their} possible association with the MeV to GeV $\gamma$-ray sources GRO~J1823$-$12 (Collmar \cite{Collmar:1}) and
3EG~1824$-$1514 (Paredes \etal \cite{Paredes:1}) suggests the presence of multi-GeV particles. 
Observations in 2004 (Aharonian \etal \cite{LS5039_HESS}) ($\sim$11~hrs) with HESS established a new 
VHE $\gamma$-ray source, HESS~J1826$-$148, within 30 arcsec of the radio position of LS~5039, revealing for the first time 
that LS~5039, and hence XRBs, are capable of multi-TeV (10$^{12}$~eV) particle acceleration. The limited statistics did not
allow for detailed timing or variability analyses. 
We note that evidence for variability at VHE $\gamma$-ray energies
has recently been unveiled (Albert \etal \cite{MAGIC_LSI}) in a similar type of binary system, LS~I~$+61^\circ 303$.
Here we report on new, deeper HESS observations of LS~5039 at TeV $\gamma$-ray energies, revealing that its VHE $\gamma$-ray emission is modulated by 
the orbital motion of the compact object around its massive stellar companion.

\section{Observations}

The observations were taken with HESS (Aharonian \etal \cite{HESS_Crab}),
an array of four identical Atmospheric Cherenkov Telescopes (ACT) located in the Southern Hemisphere 
(Namibia, 1800~m a.s.l), and is sensitive to $\gamma$-rays above 0.1~TeV. 
The 2004 HESS observations (Aharonian \etal \cite{LS5039_HESS}) have been followed up with a deeper campaign in 2005. After data quality selection, 
the total dataset comprises 160 runs (or pointings) representing 69.2~hours observations from both 2004 and 2005. Data were analysed, employing 
two separate calibration procedures (Aharonian \etal \cite{Aharonian:2}) and several background rejection and direction reconstruction methods.
The results presented here are based on the combination of a semi-analytical shower model and a parametrisation based on the moment 
method of Hillas to yield the combined likelihood of the event being initiated by a $\gamma$-ray primary (de~Naurois \etal \cite{Mathieu:1}). 
As we show later, a pure Hillas-based analysis, described in Aharonian \etal (\cite{HESS_Crab}), also yielded consistent results.

\section{Results}

A total of 1960 $\gamma$-ray events (with an excess significance above the background exceeding +40$\sigma$) 
within 0.1$^\circ$ of the VLBA radio position of 
LS~5039 (Rib\'o \etal \cite{Ribo:1}) were found. The best-fit position (in Galactic coordinates) 
is $l=16.879^\circ$, $b = -1.285^\circ$ with statistical and systematic uncertainties of $\pm$12 and $\pm$20 arcsec, respectively,
which is consistent with the VLBA position within the 1$\sigma$ statistical uncertainty.  

\subsection{Timing Analysis}

\begin{figure}[h]
  \centering
    \includegraphics[width=9cm]{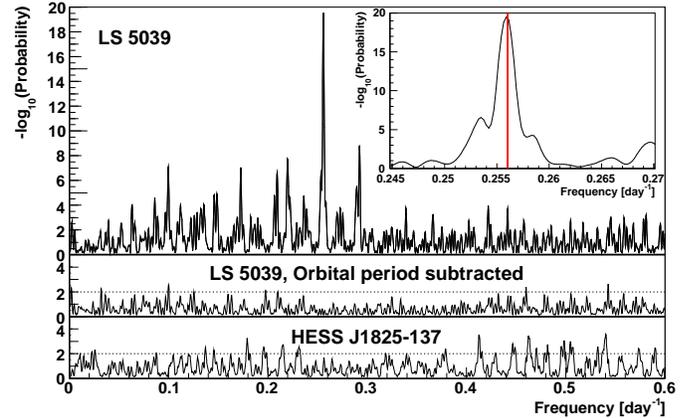}
    \vspace*{-2mm}
    \caption{{\bf Top:} Lomb-Scargle (LS) periodogram of the VHE runwise flux for LS~5039 (chance probability to obtain the LS power vs. frequency). 
      Inset: zoom around the highest
      peak (pre-trial probability $\sim$10$^{-20}$), which corresponds to a period of 3.9078$\pm$0.0015~days, compatible with the ephemeris 
      value of 3.90603$\pm 0.00017$ days (vertical red line at 0.2560~days$^{-1}$ on the inset) from Casares \etal (\cite{Casares:1}).  
      The post-trial chance probability of the orbital period peak is found to be less
      than 10$^{-15}$ (see Fig.~\ref{fig:trials}). 
      {\bf Middle:} LS periodogram of the same data after subtraction of
      a pure sinusoidal component at the orbital period of 3.90603 days. 
      The orbital frequency peak has been removed as expected, as well as significant satellite peaks (see text).
      {\bf Bottom:} LS periodogram of the HESS source HESS~J1825$-$137 observed simultaneously in the same field of view. The middle and bottom 
      panel results are consistent with that expected of white noise over the range of frequencies sampled. The dotted lines correspond 
      to a 10$^{-2}$ pre-trial chance probability.}
    \label{fig:lomb}
\end{figure}

\begin{figure}[h]
  \centering
    \includegraphics[width=9cm]{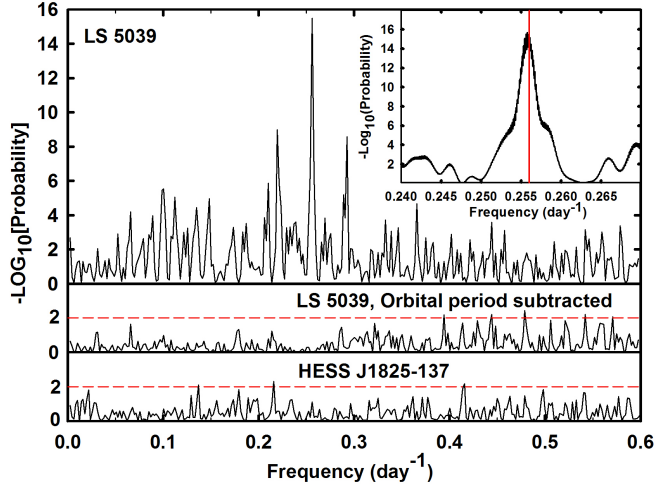}
    \vspace*{-2mm}
    \caption{{\bf Top:} Normalised Rayleigh Statistic (deJager \cite{Rayleigh:1}) periodogram calculated from run-wise 
      HESS fluxes for LS~5039. 
      The {\bf middle} and {\bf bottom} panels depict the NRS after subtraction of the orbital period and for HESS~J1825$-$137, respectively 
      (As for the Lomb-Scargle test in Fig.~\ref{fig:lomb}).}
  \label{fig:nrs}
\end{figure}

A search for periodicity, by decomposing the runwise VHE $\gamma$-ray flux at energies $>$ 1~TeV into its frequency components, 
was carried out using the Lomb-Scargle Test (Scargle \cite{Scargle:1}), (Fig.~\ref{fig:lomb}) and Normalised Rayleigh Statistic (NRS) 
(de~Jager \cite{Rayleigh:1}) (Fig.~\ref{fig:nrs}) 
which are appropriate for unevenly sampled datasets typical of those taken by HESS. 
The 2005 observations were taken over a wide range of zenith angles yielding a varying energy threshold in the range 0.2 to $\sim$1~TeV.
To reduce adverse affects of this varying threshold in our timing analysis, we used all events and extracted the flux normalisation above 1~TeV 
assuming an average photon power-law index derived from all data ($\Gamma=2.23$ for $dN/dE \sim E^{-\Gamma}$). 
Although as we see later the photon spectral index was found to vary within the orbital period, the average index assumption in this method contributes 
only a small error on the derived flux above 1~TeV.

\begin{figure}[h]
  \centering
  \includegraphics[width=9cm]{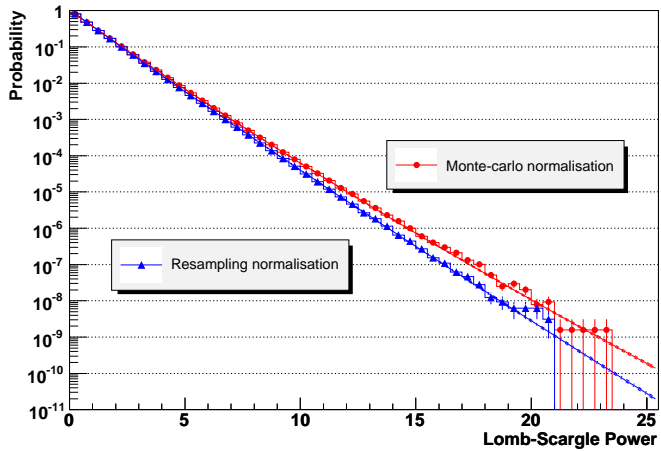}
  \vspace*{-2mm}
  \caption{Density function of the chance probability of the Lomb-Scargle power determined by Monte-Carlo
    and resampling methods after subtraction of the orbital period sinusoid.
    The expected exponential density functions are also indicated (solid lines). In the Monte-Carlo method, $\sim 10^6$ 
    random time series were generated to produce a distribution of the highest Lomb-Scargle power. The highest power obtained after these 
    iterations reaches just above 20, well below the power of 62 obtained from the unshuffled lightcurve at the orbital period. 
    An extrapolation of the curves is therefore used to estimate the chance probability for powers above 20.}
  \label{fig:trials}
\end{figure}

An obvious peak in the Lomb-Scargle 
periodogram occurs at the period 3.9078$\pm$0.0015~days (similarly observed in the NRS test), quite 
consistent with the orbital period determined by Casares \etal (\cite{Casares:1}) (3.90603$\pm 0.00017$ days) from radial velocity measurements
of the stellar companion. The error in this measurement was estimated from Monte-Carlo-simulated time series containing a sinusoid above a
random background.
The peak is highly significant, with a post-trial probability of less than 10$^{-15}$
that it results from a statistical fluctuation. 
This chance probability was estimated via Monte-Carlo simulation of random fluxes and also random re-sampling of fluxes 
(Fig.~\ref{fig:trials}). 
In Fig.~\ref{fig:lomb} (middle panel) we also show the effect of subtracting the orbital period, which removes numerous satellite peaks that are
beat periods of the orbital period with the various gaps present in the HESS dataset (1-day, 28-day moon cycle, 365.25-day annual), that is,
rational fractions of beat periods added to the orbital period. Fig.~\ref{fig:lomb} (bottom panel) also includes results obtained on the neighbouring
VHE $\gamma$-ray source HESS~J1825$-$137 (Aharonian \etal \cite{HESSJ1825}), which is in the same field of view (FoV) as LS~5039 and therefore 
observed simultaneously.  The HESS~J1825-137 periodogram does not show statistically significant
peaks, demonstrating that the significant peak is genuinely associated with LS~5039.

\begin{figure}[h]
  \centering
  \includegraphics[width=8.5cm]{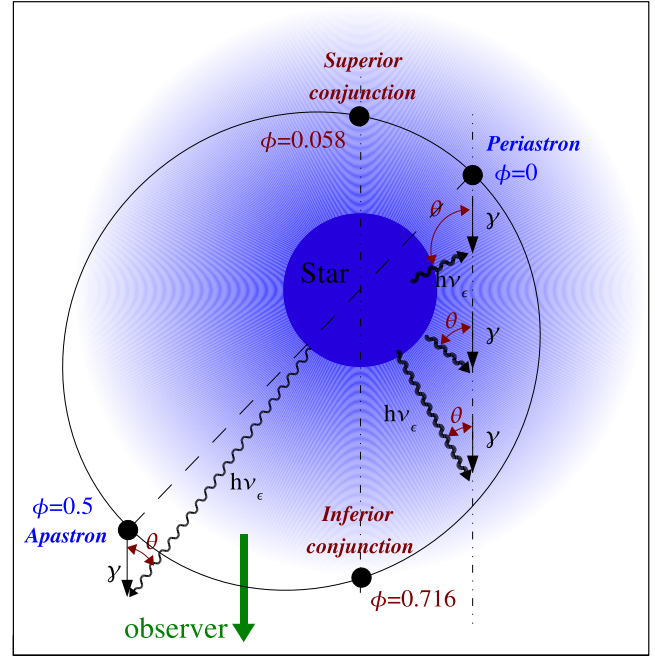}
  \caption{The orbital geometry (Casares \etal \cite{Casares:1}) viewed from directly above LS~5039. 
    Shown are: phases ($\phi$) of minimum (periastron) and maximum (apastron) separation between the two components; epochs of superior and inferior 
    conjunctions of the compact object 
    representing phases of co-aligment along our line-of-sight of the compact object and stellar companion. The orbit is
    actually inclined at an angle in the range $13^\circ < i < 64^\circ$ with respect to the view above. 
    VHE $\gamma$-rays (straight black lines with arrows) can be absorbed by optical photons of energy $h\nu_\epsilon$, when their scattering
    angle $\theta$ exceeds zero.}
  \label{fig:geometry}
\end{figure}

The ephemeris of Casares \etal (\cite{Casares:1}), determined from Doppler-shifted optical lines (observed in 2002 and 2003),
shows the binary makeup  (Fig.~\ref{fig:geometry}) of LS~5039 as comprising a compact object of mass $>1.38$~M$_\odot$
\footnote{A black hole of mass $3.7^{+1.3}_{-1.0}$~M$_\odot$ for the compact object 
was derived by Casares \etal (\cite{Casares:1}) under the assumption of pseudo-synchronisation of the binary components},
in an eccentric $e=0.35$ orbit around a stellar companion of mass $\sim$20~M$_\odot$ (with bolometric
luminosity $L_* \sim 10^{39}$~erg~s$^{-1}$).
The separation (centre-to-centre) between these two components varies between 2.2$R_*$ at periastron
($\phi=0.0$ with reference epoch $T_0$(HJD-2400000.5)=51942.59) to 4.5$R_*$  at apastron ($\phi=0.5$),
for a stellar radius $R_*=7\times10^{11}$~cm.
A range on the system inclination angle of $13^\circ < i < 64^\circ$ is inferred from the binary mass function, the companion
rotation velocity, 
the lack of X-ray eclipses (which assumes that the X-ray emission occurs very close to the
compact object) and lack of Roche lobe overflow. 

\begin{figure}[h]
  \centering
  \includegraphics[width=9cm]{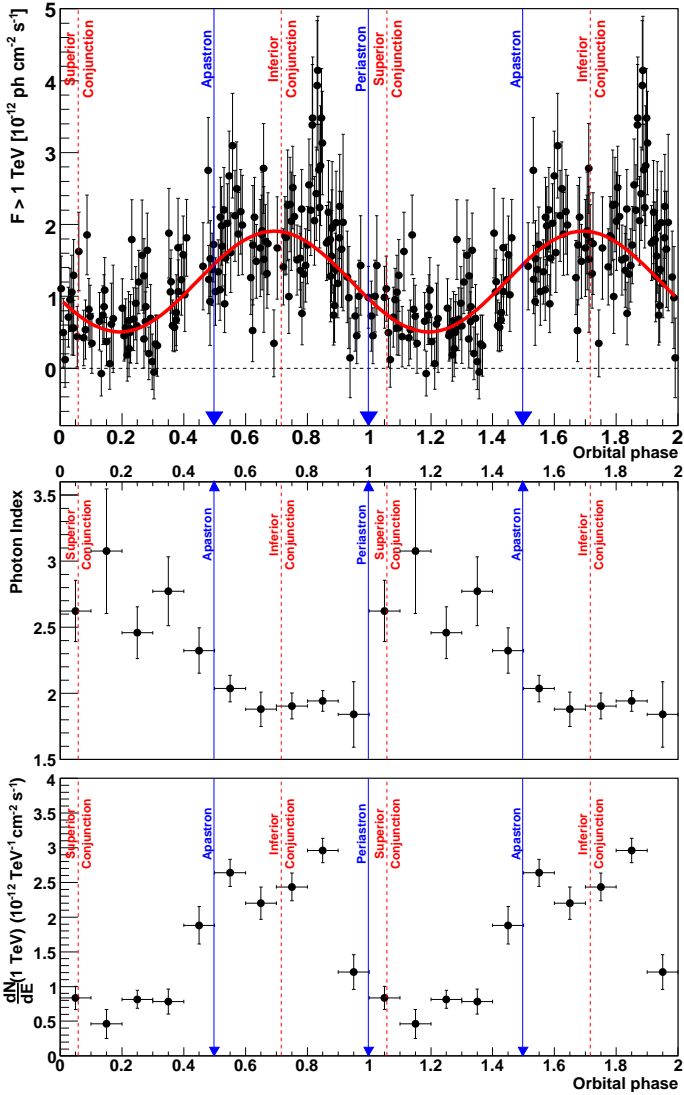}
  \caption{{\bf Top:} Integral $\gamma$-ray flux ($F>1$~TeV) lightcurve (phasogram) of LS~5039 from HESS data (2004 to 2005) 
    on a run-by-run basis folded with the orbital ephemeris of Casares \etal (\cite{Casares:1}).  Each run is $\sim$28~minutes.
    Two full phase ($\phi$) periods are shown for clarity. 
    The blue solid arrows depict periastron and apastron.
    The thin red dashed lines represent the superior and inferior conjunctions of the compact object, 
    and the thick red dashed line depicts the Lomb-Scargle Sine coefficients for the
    period giving the highest Lomb-Scargle power. This coefficient is subtracted from the light curve in Fig~\ref{fig:lomb} middle panel.
    {\bf Middle:} Fitted pure power-law photon index (for energies 0.2 to 5~TeV) vs. phase interval of width $\Delta \phi=0.1$. 
    Because of low statistics in each bin, more complicated functions such as a power-law with
    exponential cutoff provide a no better than a pure power-law.
    {\bf Bottom:} Power-law normalisation (at 1~TeV) vs. phase interval of width  $\Delta \phi=0.1$.}
  \label{fig:lightcurves}
\end{figure}

The phasogram (Fig.~\ref{fig:lightcurves} top) of integral fluxes at energies $E>1$~TeV 
vs. orbital phase ($\phi$) obtained on a run-by-run basis (one data run is $\sim$28~minutes) clearly indicates that the 
bulk of the VHE $\gamma$-ray 
emission is confined to roughly half of the orbital period, covering the phase interval $\phi \sim$0.45 to 0.9. 
The VHE flux maximum appears to lag somewhat behind the apastron epoch, and aligns better with {\it inferior conjunction}  ($\phi=0.716$) 
of the compact object. Inferior conjunction of the compact object occurs when it is lined up along our line-of-sight
in front of the stellar companion. The VHE flux minimum occurs at 
a phase $\phi \sim 0.2$, slightly further along the orbit than {\it superior conjunction} ($\phi=0.058$), 
which is when the compact object is lined up behind the stellar companion. 
Note that the inclination upper limit $i<64^\circ$
implies that direct views of both compact object and stellar companion are always available. 
We define here two broad phase intervals for further study: \textbf{INFC} ($0.45<\phi\leq0.9$) encompassing inferior conjunction, and \textbf{SUPC}
($\phi \leq 0.45$ and $\phi>0.9$) likewise for superior conjunction. 
The phase error ($\Delta \phi = (T \Delta P)/P^2 = 0.01$) due to uncertainties in the period measurement $\Delta P = 0.00017$ from Casares \etal
(\cite{Casares:1}), the dataset length $T\sim 1000$~days and $P=3.9$~days, appears to be negligible. Nevertheless, further near-future 
optical line observations bracketing ours at VHE $\gamma$-ray observations would be desirable to check for the presence of systematic drifts 
in the orbital period. 

\subsection{Phase-Resolved Energy Spectra}

\begin{figure}[h]
  \centering
  \includegraphics[width=9cm]{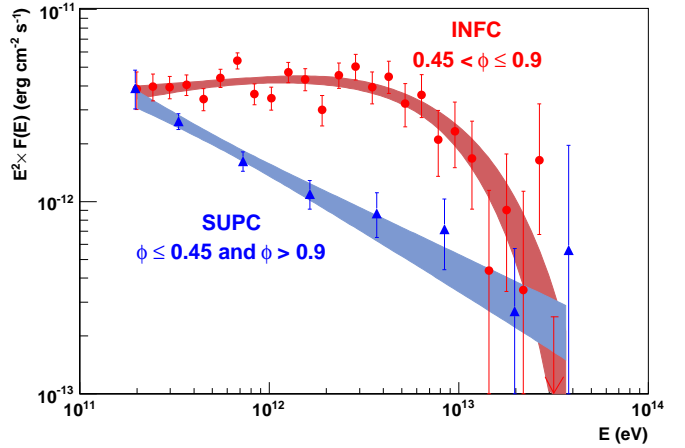}
  \vspace*{-2mm}
  \caption{Very high energy $\gamma$-ray spectra of LS~5039 for two broad orbital phase intervals (defined in the text):
    \textbf{INFC} $0.45<\phi\leq0.9$ (red circles); \textbf{SUPC} $\phi \leq 0.45$ and $\phi>0.9$ (blue triangles).
    The shaded regions represent the 1$\sigma$ confidence bands on the fitted functions (Tab.~\ref{tab:spec}).
    Both spectra are mutually incompatible with the probability that the same spectral shape would fit both simultaneously
    being $\sim2\times10^{-6}$. A clear spectral hardening in the region 0.3 to $\sim 20$~TeV is noticed for the \textbf{INFC} phase
    interval.}
    \label{fig:spectra}
\end{figure}

The energy spectrum of the VHE $\gamma$-ray emission, and in particular how it might vary with orbital phase, is an important diagnostic tool. 
The differential photon energy spectrum (see Fig.~\ref{fig:spectra}) (0.2 to 10.0~TeV) for \textbf{INFC} is consistent with a 
hard power-law where $\Gamma=1.85\pm0.06_{\mathrm{stat}} \pm 0.1_{\mathrm{syst}}$ with exponential cutoff at $E_o=8.7\pm2.0$~TeV
(for fitted function $dN/dE \sim E^{-\Gamma}\, \exp(-E/E_o)$). In contrast, the spectrum for \textbf{SUPC} is consistent 
with a relatively steep ($\Gamma=2.53\pm0.07_{\mathrm{stat}} \pm 0.1_{\mathrm{syst}}$) pure power-law (0.2 to 10 TeV) 
(see Fig.~\ref{fig:spectra}). 
The spectra from these phase intervals are mutually incompatible, with the probability that the same spectral shape would fit both simultaneously
being $\sim2\times10^{-6}$.
Fitting a pure power-law (which is statistically sufficient at present) to narrower phase intervals of width $\Delta \phi = 0.1$, 
and restricting the fit to energies $E\leq 5$~TeV to reduce
the effect of any cutoff, also demonstrates that a harder spectrum occurs when the flux is higher 
(Fig.~\ref{fig:lightcurves} middle and bottom panels).
Notably, the VHE flux at $E\sim 0.2$~TeV appears to be quite stable over phases and the 
strongest modulation occurs at a few TeV.

\begin{figure}[h]
  \centering
    \includegraphics[width=9cm]{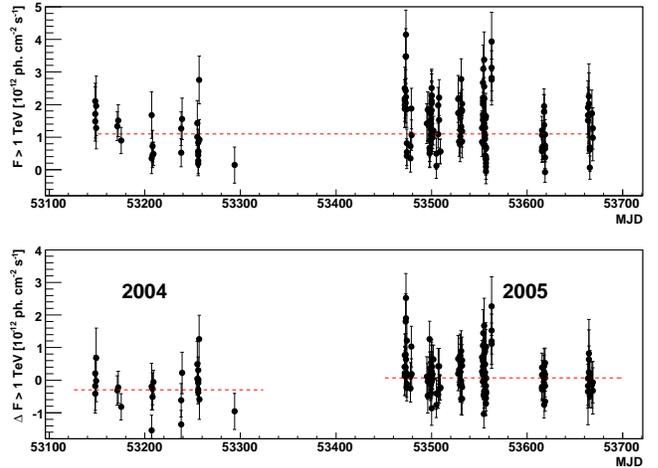}
    \vspace*{-2mm}
  \caption{{\bf Top:} Integral flux ($E>1$~TeV) vs. time (MJD) for LS~5039 on a run-by-run basis. {\bf Bottom:} After
    subtraction of the orbital period of 3.9063 days (this is achieved by subtraction of the Lomb-Scargle coefficients for the selected
    period). The average flux (dashed lines) for the post-subtracted light-curve is consistent with a steady source with a chance probability of 
    4$\times 10^{-2}$. In addition, the average flux levels determined exclusively for 2004 and 2005 data differ at a 2$\sigma$ level 
    (statistical errors) but are consistent when factoring in systematic errors of $\sim$10\% on a run-by-run basis.} 
  \label{fig:subtracted}
\end{figure}

\begin{figure}[h]
  \centering
    \includegraphics[width=9cm]{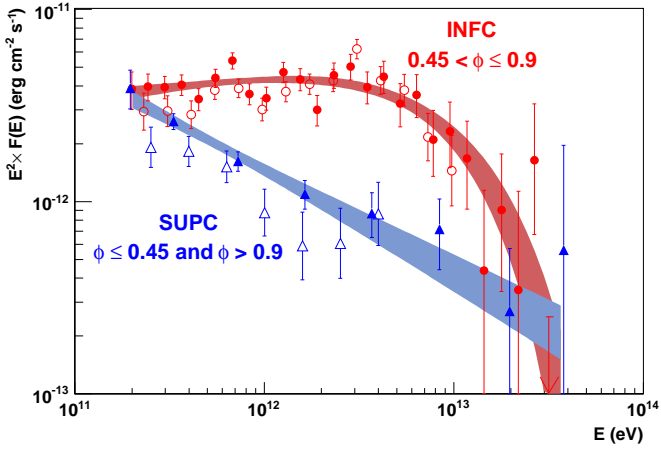}
    \vspace*{-2mm}
  \caption{Energy spectra of LS~5039 separated into the same broad phase intervals as for Fig.~\ref{fig:spectra}. 
     \textbf{INFC} $0.45<\phi\leq0.9$ (red circles); \textbf{SUPC} $\phi \leq 0.45$ and $\phi>0.9$ (blue triangles),
    comparing results from the semi-analytical Model+Hillas (filled markers) and Hillas-only (open markers) analyses 
    (Aharonian \etal \cite{HESS_Crab}), respectively. 
    The Hillas-only analysis has made use of an independent calibration chain.}
  \label{fig:hillas_model}
\end{figure}

We found no evidence for long-term secular variations in VHE flux on a yearly scale independent of the orbital modulation (Fig.~\ref{fig:subtracted}). 
The orbital modulation represents a VHE $\gamma$-ray luminosity (0.2 to 10~TeV; at 2.5~kpc) variation 
between 4 to 10 $\times 10^{33}$~erg~s$^{-1}$.
Spectral fits and other numerical results are summarised in Tab.~\ref{tab:spec}.
\begin{table*}[t]
 \caption{Photon energy spectra and luminosity (0.2 to 10~TeV) of the VHE $\gamma$-ray emission of LS~5039 for different orbital phase intervals.
   The broad phase intervals \textbf{INFC} and \textbf{SUPC} encompass the inferior and superior conjunction epochs.
   The orbital phase $\phi$ is calculated from the ephemeris of Casares \etal (\cite{Casares:1}). The best-fit 
   function is indicated for each phase interval. The errors quoted are statistical with systematic errors in the 10 to 15\% range.}
 \label{tab:spec}
 \centering
 \begin{tabular}{c|cccc} \hline \hline
   Orbital Phase Interval & $N$               & $\Gamma$ & $E_o$ & Luminosity$^\dagger$ \\
   &  $\times 10^{-12}$ [ph~cm$^{-2}$s$^{-1}$TeV$^{-1}$] & & [TeV] & [erg~s$^{-1}$]\\ \hline 
   \multicolumn{5}{c}{Fit function:  \fbox{$dN/dE = N E^{-\Gamma} \exp(-E/E_o)$}} \\ \hline
   \textbf{INFC} ($0.45<\phi\leq0.9$) & $2.28 \pm0.10$ & $1.85\pm0.06$ & 8.7$\pm$2.0 & 1.1$\times 10^{34}$\\
   \textbf{SUPC} ($\phi \leq 0.45$) and $\phi>0.9$  & $0.91 \pm0.07$  & $2.53\pm0.07$ & - & 4.2$\times 10^{33}$\\
   All Phases (time averaged)  & $1.85 \pm0.06$  & $2.06\pm0.05$ & 13.0$\pm$4.1 & 7.8$\times 10^{33}$\\ \hline
   \multicolumn{5}{c}{Power-Law fit for energies $E\leq$5~TeV (narrow phase intervals)}\\ \hline
   0.0--0.1 & 0.84 $\pm$ 0.17 & 2.62 $\pm$ 0.23 & & \\
   0.1--0.2 & 0.46 $\pm$ 0.21 & 3.08 $\pm$ 0.47 & & \\
   0.2--0.3 & 0.81 $\pm$ 0.13 & 2.46 $\pm$ 0.20 & & \\
   0.3--0.4 & 0.78 $\pm$ 0.18 & 2.77 $\pm$ 0.60 & & \\
   0.4--0.5 & 1.88 $\pm$ 0.27 & 2.32 $\pm$ 0.17 & & \\
   0.5--0.6 & 2.64 $\pm$ 0.20 & 2.04 $\pm$ 0.10 & & \\
   0.6--0.7 & 2.20 $\pm$ 0.23 & 1.88 $\pm$ 0.13 & & \\
   0.7--0.8 & 2.44 $\pm$ 0.20 & 1.90 $\pm$ 0.10 & & \\
   0.8--0.9 & 2.96 $\pm$ 0.17 & 1.94 $\pm$ 0.08 & & \\
   0.9--1.0 & 1.21 $\pm$ 0.25 & 1.84 $\pm$ 0.25 & & \\ \hline \hline
   \multicolumn{5}{l}{\scriptsize $\dagger$ At 2.5~kpc distance.}
 \end{tabular}
\end{table*}
Finally, in Fig.~\ref{fig:hillas_model} we show that the spectral results 
obtained from a pure Hillas-based analysis using a separate calibration procedure are consistent with results from the Semi-analytical Model+Hillas 
results within statistical errors.

\section{Discussion}

The basic paradigm of VHE $\gamma$-ray production requires the presence of particles accelerated to multi-TeV energies and a target 
comprising photons and/or matter of sufficient density.
In microquasars, particle acceleration could take place directly inside and along the jet, out to parsec-scale 
distances, and also at jet termination regions due to interaction with ambient matter (Heinz \& Sunyaev \cite{Heinz:1}). A non-jet scenario based on
acceleration in shocks created by the interaction of a pulsar wind with the wind of the stellar companion has also
been suggested (Maraschi and Treves \cite{Maraschi:1}, Tavani \etal \cite{Tavani:1}, Dubus \cite{Dubus:2}). 
The nature of the parent particles responsible for VHE $\gamma$-ray emission is under theoretical discussion, with both accelerated electron 
(Aharonian \& Atoyan \cite{AharAtoy:1}, Paredes \etal \cite{ParedesBB}) and hadron (Distefano \etal \cite{Distefano:1}, 
Romero \etal \cite{Romero:1}) scenarios proposed.
Observationally, electrons (eg. Corbel \etal \cite{Corbel:1}, Angelini \etal \cite{Angelini:1}) and hadrons (Margon \cite{Margon:1}) are both 
known to be present inside jets.

The orbital modulation in LS~5039, with a peak flux around inferior conjunction, minimum flux around superior conjunction, 
and hardening of energy spectra, provides new information about the physical processes in microquasars. {\bf Our results can be compared with those
at X-ray energies (3$-$30~keV), where interestingly, a spectral hardening with flux is also observed, as well as an indication for higher fluxes at $\phi \sim 0.8$ in 
the phase-resolved light curve (Bosch-Ramon \etal \cite{Valenti:2}).}  
The VHE $\gamma$-ray modulation is an unambiguous sign 
that periodic changes in the VHE $\gamma$-ray absorption and/or production processes are occurring, and we discuss briefly how these could arise
along with issues concerning the location and size of the $\gamma$-ray production region.

\subsection{Gamma-Ray Absorption}
VHE $\gamma$-rays produced close enough to the stellar companion will unavoidably 
suffer severe absorption via pair production (e$^+$e$^-$) on its intense optical photon field.
The cross-section for pair production is dependent on the angle $\theta$ between the VHE $\gamma$-ray and optical photons,
and has an energy threshold varying as $\mathrm{1/(1-\cos\theta)}$.
The level of absorption therefore depends on alignment between 
the VHE $\gamma$-ray production region, the star, and the observer, leading to orbital modulation of the 
VHE $\gamma$-ray flux  (Protheroe and Stanev \cite{Protheroe:1}, Moskalenko \cite{Moskalenko:1}, B\"ottcher and Dermer \cite{Boettcher:1},
Dubus \cite{Dubus:1}).
For the orbital geometry of LS~5039, in which apastron and periastron are near the inferior and superior conjunction epochs, 
the absorption effect provides flux maxima at the phase of inferior conjunction, when $\gamma$
photons are emitted towards the observer parallel to stellar photons ($\mathrm{\cos\theta=1}$),
and flux minima at superior conjunction (Dubus \cite{Dubus:1}). This picture 
agrees well with the observed phasogram (Fig.~\ref{fig:lightcurves}), suggesting that absorption plays an important role. 
An additional key expectation is that the strongest absorption, and hence modulation, should occur in the energy range $E\sim 0.2$ to 2~TeV 
(Dubus \cite{Dubus:1}).  
However, the observations show that the flux at $\sim 0.2$~TeV appears quite stable, suggesting that additional processes must be considered
to explain the spectral modulation.

\subsection{Gamma-Ray Production}
VHE $\gamma$-ray emission can be produced by accelerated electrons through the inverse-Compton
(IC) scattering of stellar photons of the companion star, and/or from accelerated hadrons through their
interaction with surrounding photons and particles. 
The efficiency of VHE $\gamma$-ray production under this scenario will peak around periastron ($\phi=0.0$), 
reflecting the minimal separation between particle acceleration sites and targets, and higher target photon densities. 
The observed phasogram is in contrast with this
behaviour. However, important influences on the energy spectrum can arise from variations of the maximum energy of accelerated electrons and 
dominance in the radiative processes (IC and/or synchrotron emission) by which they lose energy (known as cooling).
The high temperature of the companion star means that IC $\gamma$-ray production proceeds primarily in the deep Klein-Nishina regime
(where the IC cross-section is sharply reduced compared to the Thompson regime). The maximum electron energy $E_{\rm max}$ is derived by equating the 
competing acceleration and dominant radiative loss timescales. The hard VHE $\gamma$-ray spectrum ($\Gamma \leq2.5$) 
implies IC is the dominant cooling mechanism. Above some high energy $\epsilon$ (discussed shortly {\bf below}), synchrotron losses however are
expected to take over.  
For dominant IC cooling we have (Aharonian \etal \cite{Aharonian:3}) 
$E_{\rm max} \propto (B/w)^{3.3}$, where $B$ is the magnetic field and $w$ is the target photon energy density. 
Since $w \propto 1/d^2$ (for binary separation $d$), and assuming $B \propto 1/d$, $E_{\rm max}$ will increase by a factor $\sim$10 from 
periastron to apastron, 
leading to a spectral hardening around the apastron phase (for LS~5039, also near inferior conjunction).
For magnetic fields expected within the binary orbit, $B\sim 1$~G, electrons with energy above 
$\epsilon \approx 6 [(B/{\rm G})(d/R_*)]^{-1}$ TeV\footnote{The electron energy at which synchrotron and IC cooling times are equal.}
will cool via synchrotron (X-ray) radiation in preference to IC cooling.
Much stronger synchrotron losses will then produce a steepening of the VHE $\gamma$-ray spectrum 
(Moderski \etal \cite{Moderski:1}) and thus the $d$ and $B$ dependence of this changeover energy $\epsilon$ 
could also introduce a spectral hardening at apastron.
Spectral hardening could also come from the angular dependence of the IC scattering {\bf cross-section} (Khangulyan \& Aharonian \cite{Khang:1}), 
which peaks at small scattering angles (ie. around inferior conjunction).
It is also feasible to consider production via accelerated hadrons. For considerably larger magnetic fields, 
for example at the base of the 
jet where $B\sim 10^5$~G could be expected, {\bf protons can be accelerated to multi-TeV energies. In such high magnetic fields, the strong cooling of electrons 
will also render any IC components negligible (see for example Aharonian \etal \cite{Aharonian:3})}. Accelerated protons 
interacting with stellar wind particles and possibly X-ray photons \footnote{The X-ray disk luminosity, $\sim 10^{34}$~erg~s$^{-1}$, is marginally 
sufficient under this process to meet by itself that of the VHE $\gamma$-ray emission.} 
associated with an accretion disk, provide a plausible hadronic origin via the decay of neutral pions.

\subsection{Pair Cascades}
For magnetic fields $B<10$~G (Aharonian \etal \cite{Aharonian:3}), an extra complication arises from pair cascades 
initiated by absorption of first generation VHE $\gamma$-rays (Aharonian \etal \cite{Aharonian:3}, Bednarek \cite{Bednarek:1}). 
Cascades significantly increase the transparency of the source at $\sim$~TeV energies, reducing absorption effects.  
Interestingly we do observe a VHE signal (excess significance 6.1$\sigma$; 79 $\gamma$-rays) in a phase interval $\phi=0.0\pm 0.05$
where VHE $\gamma$-rays are unlikely to escape due to absorption (Dubus \cite{Dubus:1}). A cascade, or at least an additional, 
possibly unmodulated component from another process may be present.   

\subsection{$\gamma$-Ray Production Region: Location \& Size}
If the phasogram profile is the result of absorption, the $\gamma$-ray production region, or at least a large fraction of it, should be embedded within the 
stellar photosphere defined where the absorption optical depth $\tau_{\gamma \gamma}$ is $\geq$1. Such an optical depth occurs within $\sim$1~AU of the stellar companion. 
However this size constraint comes
with a caveat since absorption is unlikely to be the sole process present as argued earlier. An unmodulated component, possibly explaining the $\phi=0.0\pm 0.05$
signal could for example arise if parts of the source lay outside the stellar photosphere, or is always situated in front of the stellar companion along the line of sight. 
The size constraint is considerably smaller than the jets/outflows observed  out to 
a distance of $\sim$1000~AU from the binary system (Paredes \etal \cite{Paredes:2}), and similar in size to the smaller-scale jets/outflows 
observed out to $\sim$2~AU (Paredes \etal \cite{Paredes:1}).
The size constraint is also significantly smaller than the $\sim$0.3~pc (28~arcsec at 2.5~kpc) upper limit (1$\sigma$) on the source radius based on
the HESS angular resolution.

\section{Conclusions}

In conclusion, new observations by HESS have established orbital modulation of the VHE $\gamma$-ray flux and energy spectrum from the XRB LS~5039.
The flux vs. orbital phase profile provides the first indication for $\gamma$-ray absorption within an astrophysical source, suggesting 
that a large part of the VHE $\gamma$-ray production region lies inside the photosphere (within $\sim$1~AU) of the massive stellar companion. 
However, not all of the observed effects can be explained by absorption alone. Modulation of the energy spectrum could arise from 
changes in the maximum energies of electrons responsible for the radiation, changes in the dominant radiative mechanism, and/or scattering
angle dependence of the inverse-Compton scattering effect. A VHE $\gamma$-ray signal near $\phi=0.0$ may arise from pair-cascades, or an unmodulated
component produced outside the photosphere. 
These observations provide key information about the astrophysics associated with particle acceleration processes and subsequent 
VHE $\gamma$-ray production in XRBs. 
In particular, we are now able to begin to explore in detail the complex relationship between $\gamma$-ray absorption and production processes 
within these binary systems.

\begin{acknowledgements}
The support of the Namibian authorities and of the University of Namibia in facilitating 
the construction and operation of HESS is gratefully acknowledged, as is the support of 
the German Ministry for Education and Research (BMBF), the Max Planck Society, 
the French Ministry for Research, the CNRS-IN2P3 and the Astroparticle 
Interdisciplinary Programme of the CNRS, the UK Particle Physics and Astronomy 
Research Council (PPARC), the IPNP of Charles University, the South African Department 
of Science and Technology and National Research Foundation, and the University of Namibia. 
We appreciate the excellent work of the technical support staff in Berlin, Durham, Hamburg, 
Heidelberg, Palaiseau, Paris, Saclay, and Namibia in the construction and operation of the equipment. 
L.C., C.M., M.O., and M.T. are also affiliated with the European Associated Laboratory for Gamma-Ray 
Astronomy, jointly supported by CNRS and the Max Planck Society. 
\end{acknowledgements}


\end{document}